\documentclass{elsart}
\usepackage{graphicx}
\usepackage{color}
\usepackage{bm}
\usepackage{multirow}
\usepackage{epsfig}
\usepackage[english]{babel}

\begin{document}
\bibliographystyle{try}
\topmargin 0.1cm
\newcommand{\bc}           {\begin{center}}
\newcommand{\ec}           {\end{center}}
\newcommand{\bq}           {\begin{eqnarray}}
\newcommand{\eq}           {\end{eqnarray}}
\newcommand{\be}           {\begin{equation}}
\newcommand{\ee}           {\end{equation}}
\newcommand{\bi}           {\begin{itemize}}
\newcommand{\ei}           {\end{itemize}}

\newcounter{univ_counter}
\setcounter{univ_counter} {0} \addtocounter{univ_counter} {1}
\edef\HISKP{$^{\arabic{univ_counter}}$ }
\addtocounter{univ_counter}{1}
\edef\GATCHINA{$^{\arabic{univ_counter}}$ }
\addtocounter{univ_counter}{1}
\edef\PI{$^{\arabic{univ_counter}}$ }

\begin{frontmatter}

\title{Evidence for a negative-parity spin-doublet of nucleon resonances at 1.88\,GeV
}

\author[HISKP,GATCHINA]{A.V.~Anisovich},
\author[HISKP]{E.~Klempt},
\author[HISKP,GATCHINA]{V.A.~Nikonov},
\author[HISKP,GATCHINA]{A.V.~Sarantsev},
\author[PI]{H.~Schmieden},
\author[HISKP]{U.~Thoma}
\\

\address[HISKP]{Helmholtz-Institut f\"ur Strahlen- und Kernphysik der
Universit\"at Bonn, Germany}
\address[GATCHINA]{Petersburg
Nuclear Physics Institute, Gatchina, Russia}
\address[PI]{Physikalisches Institut der
Universit\"at Bonn, Germany}

\date{\today}


\begin{abstract}
Evidence is reported for two nucleon resonances with spin-parity
$J^P=1/2^-$ and $J^P=3/2^-$ at a mass just below 1.9\,GeV. The
evidence is derived from a coupled-channel analysis of a large
number of pion and photo-produced reactions. The two resonances are
nearly degenerate in mass with two resonances of the same spin but
positive parity. Such parity doublets are predicted in models
claiming restoration of chiral symmetry in high-mass excitations of
the nucleon. Further examples of spin parity doublets are found in
addition. Alternatively, the spin doublet can be interpreted as
member of the 56-plet expected in the third excitation band of the
nucleon. Implications for the problem of the
{\it missing resonances} are discussed.  \vspace{2mm}   
\end{abstract}
\end{frontmatter}

SU(3) symmetry was the prerequisite for the interpretation of mesons
and baryons \cite{gell-mann} as systems composed of quarks and
antiquarks or of three quarks, respectively, and is the basis of
quark models. As three-particle systems, nucleons - protons and
neutrons - are expected to exhibit a rich spectrum of rotational and
vibrational energy levels. The excitation levels of the nucleon are
extremely short-lived and decay in a variety of different decay
modes. Many states are predicted which overlap and are very
difficult to resolve. Only a fraction of the expected states has
been found experimentally; the absence of many states is called the
problem of the {\it missing resonances}. It is still unclear if the
{\it missing resonances} do not exist or if they escaped detection
due to the limitations of experiments performed so far. Most
information on the spectrum of excited nucleons is derived from
pion-nucleon ($\pi N$) elastic scattering experiments which are
incapable to identify resonances with weak coupling to $\pi N$.
Indeed, model calculations suggest that the {\it missing resonances}
do have weak $N\pi$ coupling \cite{Koniuk:1979vw}.

New experimental techniques and new data are obviously required.
Photoproduction of mesons offers distinctive advantages. The use of
photon beams and inelastic reactions avoid $\pi N$ in the entrance
and exit channel; polarized photon beams, polarized hydrogen
targets, and measurements of the polarization of outgoing baryons -
best accessible in the case of hyperon production - are important to
separate contributions with different quantum numbers. Different
final states are sensitive to different resonances; hence it is
important to combine different channels into a common analysis and
to search for new resonances in a variety of different reactions. In
this letter we present the results of a multichannel partial wave
analysis (PWA) of a large body of reactions, in particular of the
large data base which exists on hyperon production. From hyperon
production experiments we expect a high sensitivity to low-spin
resonances above - and close to - the $\Lambda K$ and $\Sigma K$
thresholds which range from 1610 to 1690\,MeV. This is an
interesting mass region since so far all established low-spin
negative-parity nucleon resonances have masses below 1700\,MeV.

A large data base was fitted within the Bonn-Gatchina multichannel
partial wave analysis. The data include nearly the complete
available data base on pion-induced reactions and of
photo-production off protons. In particular, data with single pion
or $\eta$ production, with hyperon production with recoiling charged
and neutral kaons, and photoproduction of 2$\pi^0$ and $\pi^0\eta$
are included in the analysis. Recent results are presented in two
longer papers \cite{Anisovich:2010an,Anisovich:2011ye} where
references are given to the data included and to papers where the
PWA method is fully described. We reported two possible solutions
which are both compatible with the full data base used in the
analysis. These two solutions are called BG2011-01 and BG2011-02,
respectively. Newly added here are recent data on $\gamma p\to
\Sigma^+K^0$ \cite{Ewald11}. In this letter we give a brief account
of the experimental findings and focus on possible interpretations
of the results.

Table~\ref{res-list} lists the positive-parity nucleon resonances
below 2.3\,GeV used in the analysis. Here, nucleon resonances are
characterized by the letter $N$, by their nominal mass from
\cite{Nakamura:2010zzi} or from us, by their isospin $I=\frac 12$,
and by their spin and parity $J^P$. Here, we concentrate on
resonances with negative parity. A discussion of positive-parity
nucleon resonances in the 2\,GeV mass range can be found elsewhere
\cite{Anisovich:2012}.

\begin{table}[pb]
\caption{\label{res-list}List of positive-parity nucleon resonances
used in the coupled channel analysis. The $\Delta$ excitations
quoted in \cite{Nakamura:2010zzi} are used in addition. For
resonances with a *, alternative solutions exist yielding different
mass values. These are discussed in the text.\vspace{2mm}}
\bc
\renewcommand{\arraystretch}{1.3}
\begin{tabular}{cccc} \hline\hline
$N_{1/2^+}(1440)$&$N_{1/2^+}(1710)$&$N_{3/2^+}(1720)$&$N_{5/2^+}(1680)$\\
$N_{1/2^+}(1875)$&$N_{3/2^+}(1900)$&$N_{5/2^+}(1875)^*$&$N_{7/2^+}(1990)^*$\\
$N_{1/2^+}(2100)^*$&$N_{5/2^+}(2200)^*$&$N_{9/2^+}(2220)$&\\
\hline\hline
\end{tabular}\renewcommand{\arraystretch}{1.0}\ec
\end{table}

The $I(J^P)$=$\frac12(\frac12^-)$ and $I(J^P)$=$\frac12(\frac32^-)$
partial waves are described by two-pole K-matrices, the
$I(J^P)$=$\frac12(\frac52^-)$ partial wave by one pole, with
couplings to $N\pi$, $N\eta$, $\Lambda K^+$, $\Sigma K$,
$N\,(\pi\pi)_{\rm S-wave}$, $\Delta\pi$, and one unconstrained
channel (para\-meterized as $N\rho$). Amplitudes for background
contributions are included as reggeized meson exchanges in the $t$
channel and by direct couplings from initial to final states. The
poles represent the well known resonances\\[-6ex]
\begin{eqnarray}
N_{1/2^-}(1535) & N_{3/2^-}(1520) & \label{onehalf}\\
N_{1/2^-}(1650) & N_{3/2^-}(1700) & N_{5/2^-}(1675)\label{threehalf}
\end{eqnarray}
\begin{figure*}[pt]
\bc
\begin{tabular}{cc}
\epsfig{file=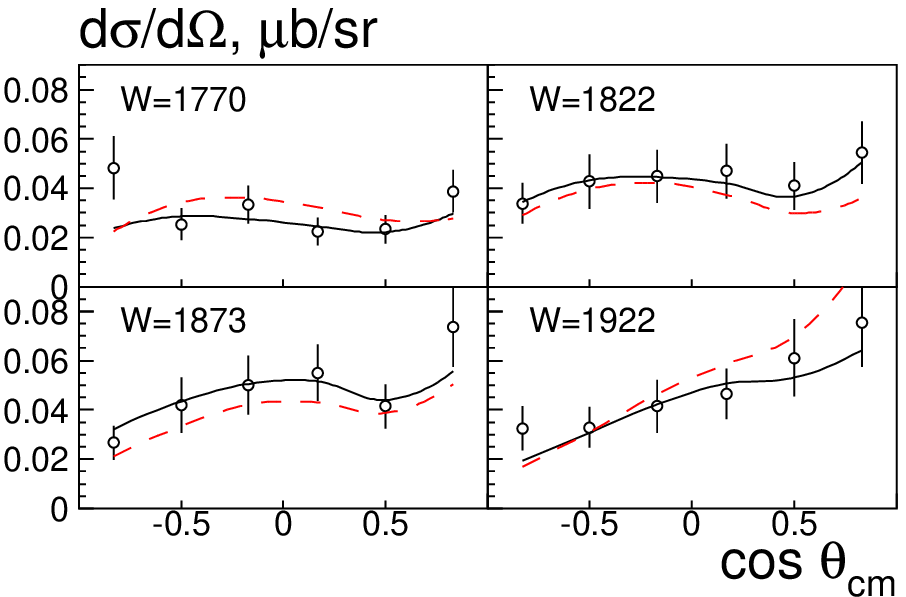,width=0.50\textwidth,clip=on}&
\hspace{-5mm}\epsfig{file=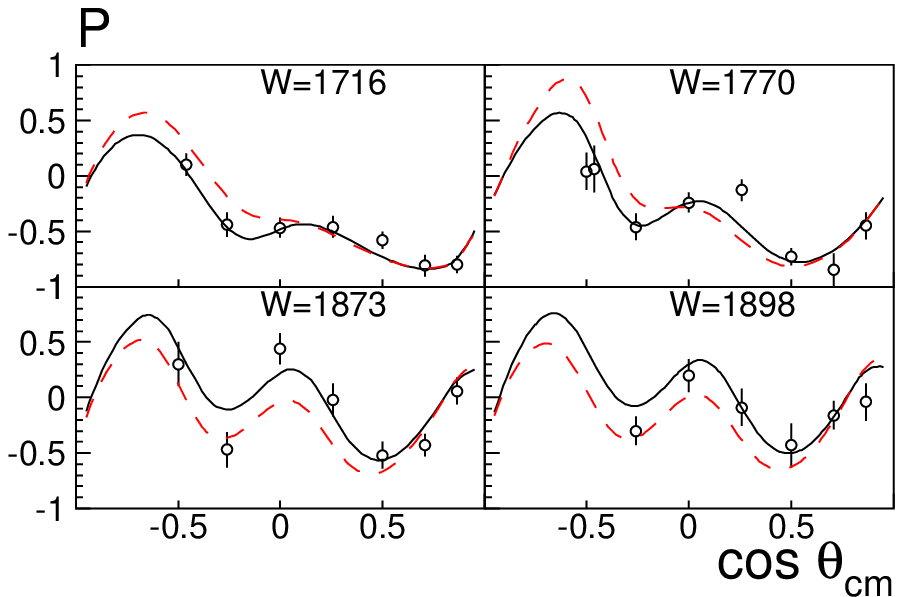,width=0.50\textwidth,clip=on}
\end{tabular}
 \ec
\caption{\label{fig1}Differential cross section for $\gamma p\to \Sigma^+ K^0$  \cite{Ewald11}
(left) and recoil asymmetry for $\gamma p\to n\pi^+$ (right) \cite{Bussey:1979ju}. The full curves
show our PWA solution BG2011-02, the dashed curves the best fit without resonant contributions
above 1.7\,GeV in the $I(J^P)$=$\frac12(\frac32^-)$ wave. }
\bc
\begin{tabular}{cc}
\epsfig{file=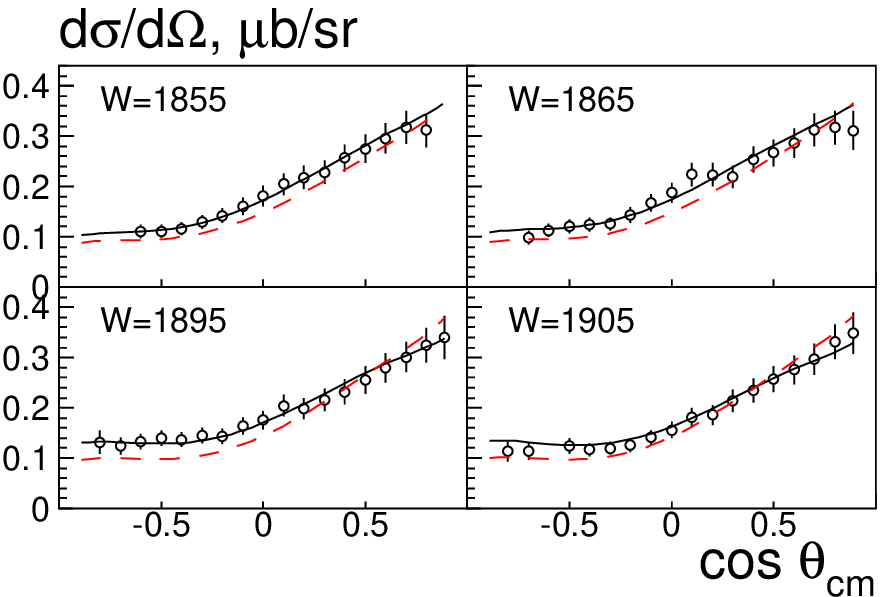,width=0.48\textwidth,clip=on}&
\epsfig{file=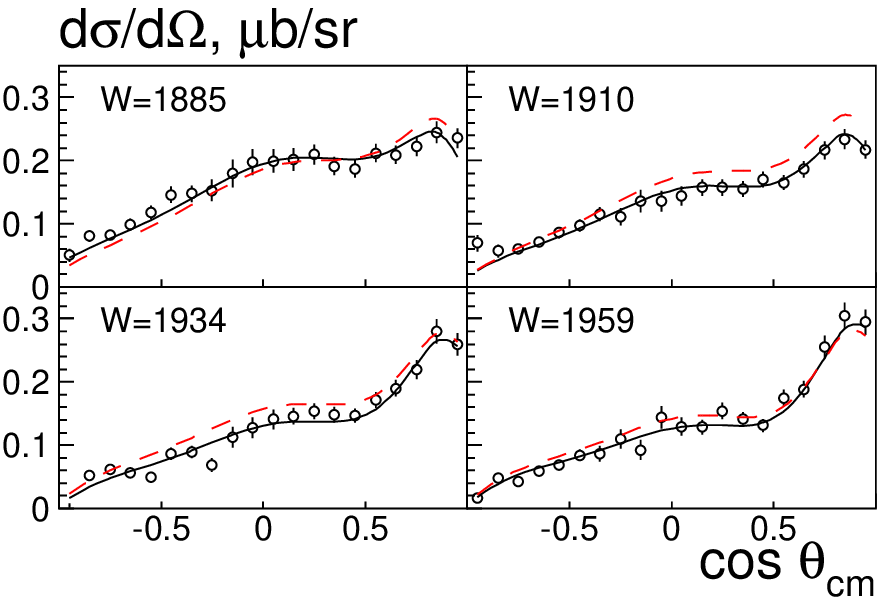,width=0.48\textwidth,clip=on}
\end{tabular}
 \ec
\caption{\label{fig2}Differential cross section for $\gamma p\to \Lambda K^+$ (left)
\cite{McCracken:2009ra} and $\gamma p\to p\eta$ \cite{Crede:2011dc} (right). The full curves show
our PWA solution BG2011-02, the dashed curves the best fit without resonant contributions above
1.7\,GeV in the $I(J^P)$=$\frac12(\frac12^-)$ wave. }
\end{figure*}
With these amplitudes, several data sets were only moderately well
described unless two further resonances were introduced, called
$N_{1/2^-}(1895)$ and $N_{3/2^-}(1875)$. In a first step, the new
resonances were represented by coupled-channel relativistic
Breit-Wigner amplitudes. With the new resonances, data and fit
agreed very well. Figures \ref{fig1} and \ref{fig2} show a few
examples, selected data on $\gamma p\to \Sigma^+ K^0_S$
\cite{Ewald11}, $\gamma p\to n\pi^+$ \cite{Bussey:1979ju}, $\gamma
p\to \Lambda K^+$ \cite{McCracken:2009ra}, and $\gamma p\to p\eta$
\cite{Crede:2011dc}. The solid lines represent the full fit, the
dashed lines in Fig. \ref{fig1} our fit when $N_{3/2^-}(1875)$ is
removed from the fit; in Fig. \ref{fig2} $N_{1/2^-}(1895)$ was
removed.

Introduction of $N_{3/2^-}(1875)$ improved the fit also to other
data which are not shown here. Significant improvements were found
in the description of the many observables in $\gamma p\to \Lambda
K^+$: in the fit to differential cross sections and recoil
polarization \cite{McCracken:2009ra}, to photon beam asymmetry
\cite{Lleres:2007tx}, target asymmetry, and to the observables
$O_{x'}$, $O_{z'}$ \cite{Lleres:2008em} and $C_x$, $C_z$
\cite{Bradford:2006ba}. The latter quantities describe,
respectively, the polarization transfer from linearly and circularly
polarized photons to the final-state hyperons. Introduction of
$N_{1/2^-}(1895)$ gave major improvements in the description of the
data on $\gamma p\to \Lambda K^+$
\cite{McCracken:2009ra,Lleres:2007tx,Lleres:2008em,Bradford:2006ba},
and for $\gamma p\to N\pi$ from different sources \cite[Table
3]{Anisovich:2010an}.

\begin{figure*}[pt]
\bc
\hspace{-5mm}\epsfig{file=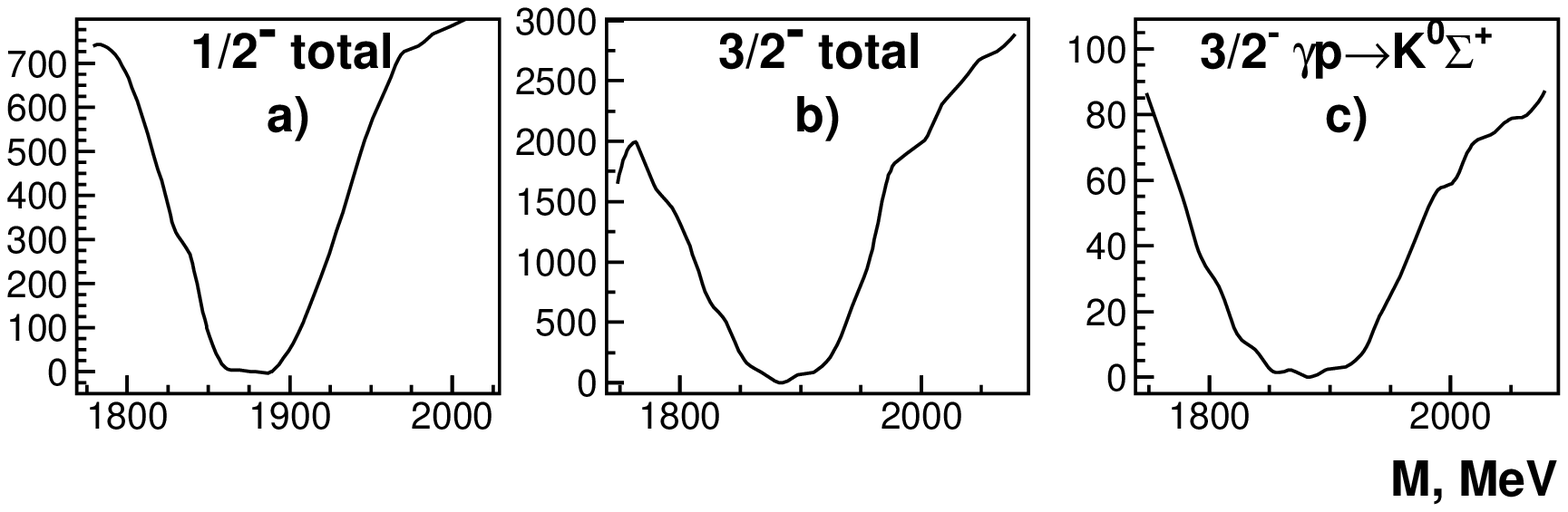,width=0.98\textwidth,clip=on}
\ec
\vspace{-4mm} \caption{\label{scan}a) Mass scan for a $N_{1/2^-}$
resonance; change of the total $\chi^2$ of the fit as a function of
the assumed mass; b, c) Mass scan for a $N_{3/2^-}$ resonance;
$\chi^2$ of the fit as a function of the assumed mass for an assumed
width of 100\,MeV. b) total $\chi^2$, c) $\chi^2$ contribution from
$\gamma p\to\Sigma^+K^0_S$ (this work).}
\end{figure*}

\begin{table}[pb]
\caption{\label{masses}Masses and widths of selected negative-parity
resonances. The second column gives the PDG \cite{Nakamura:2010zzi}
star rating, ranging from 4-star (established) to 1-star (poor
evidence). \vspace{2mm}}
\bc
\renewcommand{\arraystretch}{1.3}
\begin{tabular}{ccccc}\hline\hline
$N_{1/2^-}(1895)$: && $M_{\rm BW}=1895\pm 15$ &$\Gamma_{\rm
BW}=90^{+30}_{-15}$&\lbrack MeV \rbrack \\
$N_{3/2^-}(1875)$: && $M_{\rm BW}=1880\pm 20$ &$\Gamma_{\rm
BW}=200\pm 25$&\lbrack MeV \rbrack \\\hline
$N_{1/2^-}(2090)$: &1*& \multicolumn{2}{c}{no evidence}& \\
$N_{3/2^-}(2150)$: &2*& $M_{\rm BW}=2150\pm 60$ &$\Gamma_{\rm
BW}=330\pm 45$&\lbrack MeV \rbrack\\
$N_{5/2^-}(2060)$: &2*& $M_{\rm BW}=2060\pm 15$ &$\Gamma_{\rm
BW}=375\pm 25$&\lbrack MeV \rbrack\\
$N_{7/2^-}(2190)$: &4*& $M_{\rm BW}=2180\pm 20$ &$\Gamma_{\rm
BW}=335\pm 40$&\lbrack MeV \rbrack\\
$N_{9/2^-}(2250)$: &4*& $M_{\rm BW}=2280\pm 40$ &$\Gamma_{\rm
BW}=520\pm 50$&\lbrack MeV \rbrack\\\hline\hline
\end{tabular}
\ec\renewcommand{\arraystretch}{1.0}\end{table}
The need to introduce $N_{1/2^-}(1895)$ and $N_{3/2^-}(1875)$ can be
seen in mass scans. The mass of one of the two resonances was
stepped through the resonance region, a new fit was made with all
parameters released, except the mass of the resonance. The quality
of the fit  - expressed as $\chi^2$ as a function of the imposed
mass - was monitored. Fig.~\ref{scan}a shows a mass scan for the
$N_{1/2^-}$ resonance, Fig.~\ref{scan}b,c for $N_{3/2^-}$. The scans
show very clear and highly significant minima. Formally, the
statistical significance for $N_{1/2^-}(1895)$ corresponds to 25
standard deviations, the significance for $N_{3/2^-}(1875)$ is even
higher. The widths of the minima in Fig.~\ref{scan} reflects the
natural width of the resonance. We believe that the minima in
Fig.~\ref{scan} constitute solid evidence for the existence of these
two resonances.

In a second scan, the new resonances were included as third K-matrix
poles in the two partial waves and a search was made for higher-mass
resonances. In the $N_{3/2^-}$ wave, a clear minimum was observed at
2125\,MeV which we identify with the known two-star
$N_{3/2^-}(2200)$ \cite{Nakamura:2010zzi}. A scan for a further
$N_{1/2^-}$ resonance - known as one-star $N_{1/2^-}(2090)$
\cite{Nakamura:2010zzi} - showed no significant additional minimum.
We searched for other high-spin nucleon resonances; the results,
summarized in Table~\ref{masses}, confirm established particles.

\begin{table}[pb]
\caption{\label{csr0}Nucleon resonances as parity doublets. For an
easier comparison we give our mass values and not the nominal values
from PDG \cite{Nakamura:2010zzi}. A star$^*$ denotes values which
are not uniquely defined, a second solution exists with a different
mass. \vspace{2mm}}
{\footnotesize\renewcommand{\arraystretch}{1.3}\bc
\begin{tabular}{ccccc}\hline\hline
\hspace{-1mm}$N_{1/2^-}(1895)$&\hspace{-1mm}$N_{3/2^-}(1875)$
&\hspace{-1mm}$N_{5/2^-}(2060)$&\hspace{-1mm}$N_{7/2^-}(2190)$&\hspace{-1mm}$N_{9/2^-}(2250)$\\
\hspace{-1mm}$N_{1/2^+}(1875)$&\hspace{-1mm}$N_{3/2^+}(1900)$
&\hspace{-1mm}$N_{5/2^+}(2095)^*$&\hspace{-1mm}$N_{7/2^+}(2110)^*$&\hspace{-1mm}$N_{9/2^+}(2200)$\\\hline\hline
\end{tabular}\renewcommand{\arraystretch}{1.0}\vspace{4mm}\ec}
\end{table}
\begin{table}[pt]
\caption{\label{csr}Nucleon and $\Delta$ resonances assigned to the
third excitation shell. The masses of nucleon resonances are from
our work, most $\Delta$ resonances from \cite{Nakamura:2010zzi},
one$^*$ from \cite{Arndt:2006bf}. The two resonances
$N_{5/2^-}(2060)$ and $N_{7/2^-}(2190)$ could belong to the $S=3/2$
quartet or to a $S=1/2$. Two resonances are missing.
$N_{5/2^-}(2060)$ and $N_{7/2^-}(2190)$ may both consist of two
unresolved resonances, one belonging to the quartet, the other one
to the doublet.\vspace{2mm}}
\renewcommand{\arraystretch}{1.3}\bc
\begin{tabular}{cccccc}\hline\hline
\multirow{2}{*}{$L=1, N=1$} &S=$\frac12$&$N_{1/2^-}(1895)$&$N_{3/2^-}(1875)$& &\\
&S=$\frac32$&$\Delta_{1/2^-}(1900)$&$\Delta_{3/2^-}(1940)$&$\Delta_{5/2^-}(1930)$\\\hline
\multirow{3}{*}{$L=3, N=0$}&S=$\frac32$&$N_{3/2^-}(2150)$&\multirow{2}{*}{$N_{5/2^-}(2060)$}
&\multirow{2}{*}{$N_{7/2^-}(2190)$}&$N_{9/2^-}(2280)$\\
&S=$\frac12$&&&&\\
&S=$\frac12$&&$\Delta_{5/2^-}(2223)^*$&$\Delta_{7/2^-}(2200)$&\\\hline\hline
\end{tabular}\renewcommand{\arraystretch}{1.0}\ec
\end{table}

Evidence for the two resonances $N_{1/2^-}(1895)$ and
$N_{3/2^-}(1875)$ has been reported before. From $\pi N$ scattering,
H\"ohler {\it et al.} \cite{Hohler:1984ux} gave Breit-Wigner
parameters of $M= 1880\pm20, \Gamma =95\pm 30$\,MeV for a pole in
the $I(J^P)=\frac12(\frac12^-)$ wave. Manley {\it et al.}
\cite{Manley:1992yb} found a broad state, $M= 1928\pm59, \Gamma
=414\pm 157$\,MeV which is possibly related to the resonance
discussed here. Vrana {\it et al.} \cite{Vrana:1999nt} reported a
pole at $M_{\rm pole}= 1795, \Gamma_{\rm pole} =220$\,MeV. A third
and a forth pole in the $I(J^P)=\frac12(\frac12^-)$ wave was
suggested in \cite{Tiator:2010rp}. The third pole was given with
mass and width of $M_{\rm pole}=1733$\,MeV; $\Gamma_{\rm pole}=
180$\,MeV, and in \cite{Hadzimehmedovic:2011ua} with $M_{\rm
pole}=1745\pm80$; $\Gamma_{\rm pole}= 220\pm95$\,MeV. A forth pole
in this partial wave may have been seen by Cutkosky {\it et al.}
\cite{Cutkosky:1980rh} at $M_{\rm pole}= 2150\pm70, \Gamma_{\rm
pole} =350\pm 100$\,MeV and confirmed by Tiator {\it et al.}
\cite{Tiator:2010rp}.

In the $\frac12(\frac32^-)$ wave, Cutkosky {\it et al.}
\cite{Cutkosky:1980rh} reported two resonances, the lower mass state
at  $M_{\rm pole}= 1880\pm100, \Gamma_{\rm pole} =160\pm 80$\,MeV,
the higher mass pole at $M_{\rm pole}= 2050\pm70, \Gamma_{\rm pole}
=200\pm 80$\,MeV. A few further suggestions exist, partly supporting
the lower mass, partly the higher mass \cite{Nakamura:2010zzi}.
Based on SAPHIR data on $\gamma p\to \Lambda K^+$
\cite{Glander:2003jw}, Mart and Bennhold claimed evidence for a
$\frac12(\frac32^-)$ resonance at 1895\,MeV \cite{Mart:1999ed} which
was confirmed by us on a richer data base in
\cite{Anisovich:2005tf,Sarantsev:2005tg}, with mass and width of
$1875\pm25$ and $80\pm 20$\,MeV, respectively. The high-mass
$N_{3/2^-}$ was also seen in
\cite{Anisovich:2005tf,Sarantsev:2005tg} with $M=2166^{+25}_{-50}$;
$\Gamma=300\pm65$\,MeV and in \cite{Schumacher:2010qx} with of
$M=2100\pm20$\,MeV and $\Gamma=200\pm50$\,MeV.

We now discuss possible interpretations. Hadron resonances often
appear in parity doublets \cite{Glozman:2007ek}. Table \ref{csr0}
shows a striking consistency with this conjecture. In particular,
the new negative-parity spin-doublet $N_{1/2^-}(1895)$,
$N_{3/2^-}(1875)$ is mass degenerate with $N_{1/2^+}(1875)$ and
$N_{3/2^+}(1900)$. Also the masses of the higher-spin resonances of
opposite parity are consistent. There is, however, one caveat.
$N_{5/2^+}(2095)$, with 2000\,MeV nominal mass
\cite{Nakamura:2010zzi}, is not well defined. The $I(J^P)=\frac12
(\frac52 ^+)$ wave can be described by one resonance above
$N_{5/2^+}(1680)$. In this case, mass and width are determined to
$M,\Gamma$=($2090$$\pm20$),($450$$\pm40$)\,MeV. (In Table \ref{csr0}
we list this resonance as $N_{5/2^+}(2095)$ to avoid confusion with
$N_{3/2^-}(2090)$.) If the wave is described by three poles, the
(Breit-Wigner) mass and width of the highest pole is found at
($2190$$\pm40$),($550$$\pm100$)\,MeV, and a further pole shows up.
At present, its position cannot be defined precisely; any mass
between 1800 and 1950\,MeV gives a good description of the data.
Also for $N_{7/2^+}(1990)$ there are two solutions. In one solution,
its mass and width are determined to
($2100$$\pm15$),($260$$\pm25$)\,MeV, a mass which is consistent with
parity doubling. Solution BG2011-01 yields
($1990$$\pm10$),($180$$\pm25$)\,MeV. $N_{3/2^-}(2150)$ stands alone,
so far with no parity partner.

In quark models, baryon resonances are organized in SU(6) multiplets
combining spin and flavor according to the decomposition
$6\otimes6\otimes6=56_S\oplus70_M\oplus70_M\oplus20_A$
\cite{Hey:1982aj}. The multiplets are characterized by the SU(6)
dimensionality $D$, the leading orbital angular momentum $L$, the
shell number $\tt N$ and the parity $P$ in the form $(D,L^P_{\tt
N})$. Instead of $\tt N$, we often use the radial excitation quantum
number $N$, with ${\tt N} = L+2N$. Restricted to non-strange
baryons, the 56-plet decomposes into a spin-doublet of nucleon
resonances and a spin quartet of $\Delta$ resonances,
$56=^4\hspace{-1mm}10\oplus^2\hspace{-1mm}8$. A 70-plet is formed by
a spin quartet and a spin doublet of nucleon resonances and a spin
doublet of $\Delta$ resonances.

The two resonances $N_{1/2^-}(1895)$ and $N_{3/2^-}(1875)$ could
form a spin doublet like eq.~(\ref{onehalf}) or be members of a spin
triplet like eq.~(\ref{threehalf}). In the latter case, a close-by
resonance with $I(J^P)$=$\frac12(\frac52^-)$ should be expected. A
scan gives a minimum - with a gain in $\chi^2$ of 2500 units - at
2075\,MeV, seemingly unrelated to $N_{1/2^-}(1895)$ and
$N_{3/2^-}(1875)$. Hence we interpret these two resonances as spin
doublet. The spin doublet is not accompanied by a close-by spin
quartet (degenerate into a triplet like in eq. (\ref{threehalf});
$L=1$ and $S=3/2$ are combined to $J^P=\frac12^-, \frac32^-,
\frac52^-$). Hence the doublet must belong to a 56-plet. The
expected spin quartet of $\Delta$ resonances is degenerate to a
triplet. Indeed, such a triplet seems to exist. The Particle Data
Group \cite{Nakamura:2010zzi} lists $\Delta_{1/2^-}(1900)$,
$\Delta_{3/2^-}(1940)$, $\Delta_{5/2^-}(1930)$. These five states
and their quantum number assignment are listed in Table~\ref{csr}.
They can be assigned naturally to a 56-plet, and exhaust the
non-strange sector of this multiplet. Note that a 56-plet is
symmetric in its spin-flavor wave function. Hence the spatial wave
function must be symmetric, too, in spite of the odd angular
momentum. In three-particle systems, odd angular momenta with a
symmetric spatial wave function can indeed be constructed, except
for $L=1$ and $N=0$. With $N=2$, the resonances would belong to the
fifth excitation shell; due to their low mass, they have very likely
$N=1$. The five resonances belong to the $(56,1^-_3)$ multiplet.

In the mass range from 2000 to 2300\,MeV, four further nucleon and
two further $\Delta$ resonances are known which have negative parity
and which, in the harmonic oscillator approximation, can be assigned
to the third excitation shell. These are listed in the lower part of
Table~\ref{csr}. The $N_{9/2^-}(2280)$ resonance must have $L=3,
S=3/2$ coupling to $J^P=\frac92^-$ as dominant angular momentum
configuration; there could be a small $L=5$ component in the wave
function but resonances with $L=5$ as leading orbital angular
momentum are expected at much higher masses. With $L=3, S=3/2$
coupling to $\frac92^-$ as anchor, we expect a full quartet with
$J^P=\frac32^-$, $\frac52^-$, $\frac72^-$, $\frac92^-$. SU(6)
symmetry then demands the existence of an additional
$J^P=\frac52^-$, $\frac72^-$ doublet, i.e. six states in total.
Instead of six resonances, only four are observed here. Possibly,
the two expected resonances with $\frac52^-$ are unresolved, and
both hide in the one observed $N_{5/2^-}(2060)$. Likewise, two
resonances may hide within $N_{7/2^-}(2190)$. Thus only four
resonances instead of six are observed or observable. In the
$\Delta$ sector, the Particle Data Group lists one negative-parity
resonance in this mass range, $\Delta_{7/2^-}(2200)$, and SAID finds
$\Delta_{5/2^-}(2223)^*$ \cite{Arndt:2006bf}. These two resonances
form a natural spin doublet with $L=3, S=1/2$ coupling to
$J^P=\frac52^-$, $\frac72^-$. In SU(6), this group of nucleon and
$\Delta$ resonances can all be assigned to one multiplet
$(70,3^-_3)$. Apart from the problem that two pairs nucleon
resonances may hide in one observed spin doublet, the $(70,3^-_3)$
is completely filled.

Quark models predict six further multiplets which are completely
empty, $(56,3^-_3)$, $(20,3^-_3)$, $(70,2^-_3)$, $(70,1^-_3)$,
$(70,1^-_3)$, $(20,1^-_3)$. There is not one additional resonance
which may hint at the possibility that one of these multiplets may
be required. Some of these resonances would have noticeable
features. From the $(56,3^-_3)$ multiplet, a $\Delta$ resonance with
$J^P=\frac92^-$ is expected. A resonance with these quantum numbers
is observed, but at 2400\,MeV \cite{Nakamura:2010zzi}, too high in
mass to fall into the third excitation shell. We speculate that it
may have an additional unit of radial excitation and may belong to
the $(56,3^-_5)$ multiplet. The $(70,2^-_3)$ multiplet predicts a
quartet of states with $J^P=\frac12^-$, $\frac32^-$, $\frac52^-$,
$\frac72^-$, all with even angular momentum and odd parity. These
are just absent in the spectrum. Out of eight multiplets, six are
completely empty, two are fully equipped.

This is a remarkable observation: in two of the eight expected SU(6)
multiplets, all members seem to be identified experimentally. In
contrast, the other six multiplets remain completely empty. At
present, one thus should have to conclude that {\it missing
resonances} are not just voids which might be filled when new data
become available. It seems, instead, that whole multiplets are
unobserved and are possibly unobservable. If this conjecture should
be confirmed in future experiments and analyses, there must be a
dynamical reason which prohibits formation of certain SU(6)
multiplets.

In summary, we have reported evidence for a spin doublet of nucleon
resonances, $N_{1/2^-}(1895)$ and $N_{3/2^-}(1875)$. The spectrum of
negative parity resonances in this mass range shows remarkable
features. The resonances can be grouped, jointly with positive
parity states, into parity doublets. Within a quark-model
classification, the negative parity resonances around 2.1\,GeV can
be assigned to two multiplets while six multiplets remain completely
empty. It will be important to see whether indeed entire multiplets
are missing as opposed to individual states within multiplets. This
observation may hint to new features of intra-baryon dynamics.
\\

We would like to thank the members of SFB/TR16 for continuous
encouragement. We acknowledge support from the Deutsche
Forschungsgemeinschaft (DFG) within the SFB/ TR16 and from the
Forschungszentrum J\"ulich within the FFE program.

\end{document}